\documentclass[aip,apl,twocolumn,amsmath,amssymb]{revtex4}
\usepackage{bm}
\usepackage{amssymb}
\usepackage{amsmath}
\usepackage{latexsym}
\usepackage[dvips]{graphicx}
\usepackage{color}
\begin{document}
\author{Stanislav Stoupin}
\email{sstoupin@aps.anl.gov}
\affiliation{Advanced Photon Source, Argonne National Laboratory, Lemont, Illinois 60439, USA}

\author{Mikhail Zhernenkov}
\affiliation{National Synchrotron Light Source II, Brookhaven National Laboratory, Upton, New York, 11973, USA}

\author{Bing Shi}
\affiliation{Advanced Photon Source, Argonne National Laboratory, Lemont, Illinois 60439, USA}

\title{X-ray-induced photoemission yield for surface studies of solids beyond the photoelectron escape depth\footnote{submitted for publication in a peer-reviewed journal}}
\begin{abstract}
X-ray-induced photoemission in materials research is commonly acknowledged as a method with a probing depth limited by the escape depth of the photoelectrons. This general statement should be complemented with exceptions arising from the distribution of the X-ray wavefield in the material. Here we show that the integral hard-X-ray-induced photoemission yield is modulated by the Fresnel reflectivity of a multilayer structure with the signal originating well below the photoelectron escape depth. 
A simple electric self-detection of the integral photoemission yield and Fourier data analysis permit extraction of thicknesses of individual layers. The approach does not require detection of the reflected radiation and can be considered as a framework for non-invasive evaluation of buried layers with hard X-rays under grazing incidence. 
\end{abstract}


\maketitle



Surface structure of solids can be studied using penetrating hard X-rays without detection of the reflected radiation using grazing incidence X-ray photoemission spectroscopy (GIXPS) with angles of incidence in the vicinity of the critical angle for total external reflection. GIXPS was established by Henke~\cite{Henke72} as a method enabling determination of material constants and surface characterization. Further developments were performed \cite{Fadley74} including generalization to multilayer structures \cite{Chester93} followed by experimental effort \cite{Kawai95,Hayashi96}. The studies were mostly concentrated in the soft X-ray domain (photon energies $\lesssim$~5~keV) for applications in surface science. In spite of prior developments, applications of hard X-ray GIXPS remain limited to date~\cite{Fadley05,Fadley10,Kawai10}. Apart from the regime of grazing incidence hard X-ray photoemission spectroscopy has been successfully used to probe electronic properties of materials at depths consistent with the bulk environment (e.g., \cite{Sekiyama00,Gray12,DLFeng11}). Probing depths of about 100~$\rm \AA$ have been demonstrated~\cite{Dallera04}. 

Contrary to studies of electronic properties resolving energies of photoelectrons is not essential for probing structure/composition of a multilayer. A substantial simplification of a typical X-ray photoemission spectroscopy setup can be accomplished using self-detection of integral electric charge generated in the exterior of the studied object. This self-detection approach has been utilized in X-ray absorption spectroscopy in the hard X-ray regime (e.g., \cite{Martens78,Martens79,Erbil88}). 
For hard X-rays the detection technique can take advantage of negligible X-ray absorption in a light gas environment such as helium \cite{Kordesch84, Guo85}. Instead, helium is subject to efficient ionization by fast photoelectrons escaping the object, which provides enhancement in the quantum detection yield. The same approach is used in conversion electron M{\"{o}}ssbauer spectroscopy (e.g., \cite{Jones78}). 
Self-detection of hard-X-ray induced photoelectron yield in the grazing incidence geometry was used recently to study an X-ray mirror enclosed in a flowing helium gas~\cite{Stoupin_APL16}. It was shown that the measured integral photoelectron yield as a function of the incidence angle contains structural information
i.e., Kiessig fringes, which originate from the layered structure of the mirror. 

In this letter we show that the structural information extracted from the integral grazing-incidence yield photoemission curves is not limited by the escape depth of photoelectrons but rather is limited by the penetration depth of the X-ray wave. A bi-layer Pt-Cr system on a Si substrate was studied with a deeply buried Cr layer (116.5-$\rm \AA$-thick layer of Cr under 725-$\rm \AA$-thick layer of Pt). Detection of the layered structure using Fourier analysis of the integral photoemission yield is demonstrated at several different photon energies of the incident hard X-rays where the effective photoelectron escape depth is substantially smaller than the depth of the buried layer. 


\begin{figure}[h]
\centering\includegraphics[width=0.47\textwidth]{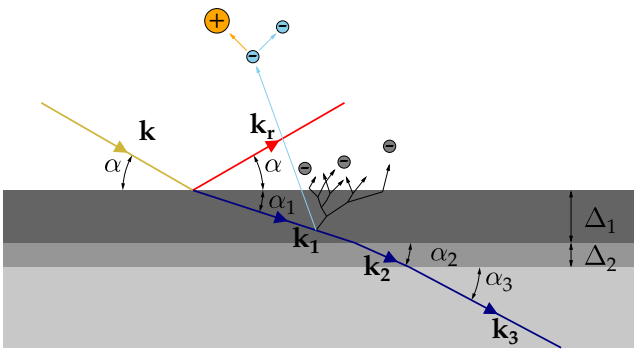}
\caption{Grazing incidence geometry illustrating the incident, the transmitted and the reflected waves. The energy flow through the layered structure generates ionization events due to X-ray photoabsorption. The escaping fast photoelectons (light blue) ionize the surrounding gas while the low-energy secondary photoelectrons (gray) do not participate in this ionization process. While the ionizing photoelectrons may escape only from a limited depth in the top layer the integral photoelectron yield is affected by the total energy flow in the multilayer and thus is sensitive to presence of buried layers.}
\label{fig:gzi}
\end{figure} 

The energy flow through a multilayer system (Fig.\ref{fig:gzi}) and thus the number of generated photoelectrons is proportional to the fraction of energy attenuated by the system, $A = 1 - R - T$, where $R$ is the net reflectivity of the system and $T$ is the net transmissivity. If the sample is thick such that the fraction of the transmitted X-rays is negligible, then $A \approx 1 - R$ represents Fresnel transmissivity of the entrance interface.
Attentuation of x rays at depth $z$ within a layer of material having a thickness $dz$ is generally described by the difference between the incoming and outgoing mean energy flow. The energy flow through the layer is represented by the real part of projection $S_z (Q,z)$ of the Poynting vector as a function of the wavevector transfer $Q = 2 k \sin{\alpha}$, where $k$ is the absolute value of the wavevector of the incident wave and $\alpha$ is the glancing angle of incidence. The normalized total X-ray attenuation per unit depth $dz$ is

\begin{equation}
s_n(Q,z) = Re \bigg\{ \frac{1}{S^0_z(Q)}\frac{d S_z(Q,z)}{dz} \bigg\} ,
\end{equation}
where $S^0_z(Q)$ is the projection of the Poynting vector of the incident wave.
The dependence on the photon energy $E_X$ of the incident wave is omitted here for clarity.

For an infinitely thick mirror the penetration depth of the wave transmitted through the entrance interface is given by 
\begin{equation}
\Lambda(Q) = \frac{1}{Im(Q_1)},
\label{eq:Lam}
\end{equation} 
where 
\begin{equation}
Q_1 = \sqrt{Q^2 - 8k^2\delta +i8k^2\beta} 
\label{eq:Q1}
\end{equation}
is the wavevector transfer in the mirror material with an index of refraction $n = 1 - \delta +i \beta$. It can be shown that for the thick mirror 
\begin{equation}
s_n(Q,z) = \frac{1}{\Lambda(Q)} T(Q) \exp[-z/\Lambda(Q)],
\label{eq:sn_tm}
\end{equation}
where $T (Q) = 1 - R(Q)$ is the Fresnel transmissivity.
If the thickness of the top layer of a multilayer X-ray mirror is such that the transmitted wave is preferentially attenuated in this layer the energy flow can be approximated with Eq.~\ref{eq:sn_tm}. 

Generation of charge carriers above the surface of the mirror is initiated with the escape of photoelectrons. An exponential factor 
$\exp[-z/L]$ (where $L$ is the effective photoelectron escape depth) can be used (e.g., \cite{Stohr_book}) to model propagation of photoelectrons towards the surface of the mirror prior to escape. Taking this factor into account, integration of Eq.~\ref{eq:sn_tm} results in the following expression for the integral electron quantum yield (yield normalized by the incident photon flux) \cite{Stoupin_APL16}.

\begin{equation}
Y (Q) = \frac{1}{2} \epsilon_q n_q G^e T(Q) \frac{L}{\Lambda(Q) + L},
\label{eq:Y}
\end{equation} 
where $\epsilon_q$ is the charge collection efficiency, $n_q$ is the charge amplification factor, and $G^e$ is a proportionality factor, which represents a correction for photoelectron energy conversion. Strictly speaking, only a fraction of the attenuated intensity of the X-rays results in generation of photoelectrons. 
This photon-energy-dependent fraction is ascribed to $G^e$ to avoid introduction of an additional factor in Eq.~\ref{eq:Y}.   

If the X-ray mirror is enclosed in a flow chamber containing light gas (e.g., He) gas impact ionization events produced by the secondary photoelectrons can be neglected (Fig.~\ref{fig:gzi}) since the energy required to produce one ion pair is $W_{g} \simeq$~40.3~eV \cite{Weiss56} while the energies of the secondary electrons do not exceed $\approx$~20~eV \cite{Henke77}. In addition, absorption cross section for hard X-rays in He is negligible compared to the ionization cross section by photoelectron impact {e.g.,\cite{Shah88}}. Thus, the electric carriers generated in the gas flow chamber originate from the photoelectric response of the mirror material. It is convenient to ascribe the number of charge carriers $n_q$ generated by a single photoelectron in the gas flow chamber to the ratio of the maximum photoelectron energy $E_{pe} \simeq E_X$ and the ion pair production energy $W_{g}$, $n_q = 2 E_X/W_g$.

Remarkably, the integral electron yield Eq.~\ref{eq:Y} is represented by the photon-electron attenuation factor $L/(\Lambda(Q) + L)$ modulated by the Fresnel transmissivity. Precise derivation of the photon-electron attenuation factor for any given material requires modelling of the photoemission processes from various atomic sub-shells and integration of the resulting photoemission cross-sections using the geometry of the photoemission detector. 
Such rigorous approach could be based on the existing theoretical developments in X-ray photoemission spectroscopy pertaining to 
X-ray optical effects \cite{Yang13}. 
It should be noted that the photon-electron attenuation factor is a slow varying function of $Q$ (or the angle of incidence $\alpha$). Thus, subtraction of a smooth function, which agrees with the overall shape of the experimental curve should isolate the modulating signal, which contains Kiessig fringes. Structural information can be extracted from the result of the subtraction using Fourier transform similarly to Fourier analysis of the interference structure in X-ray specular reflection \cite{Sakurai92}. This strategy is illustrated below applied to a bi-layer X-ray mirror. 
%
Fresnel reflection coefficients of a multilayer system are described by the Parratt's recursive relation \cite{Parratt54,Hau-Riege_book,Windt98}:
\begin{equation}
r_i = \frac{r_{ij} + r_j e^{i Q_i \Delta_i}}{1 + r_{ij}r_j e^{i Q_i \Delta_i}}.
\label{eq:r_i}
\end{equation} 
In our notation $\Delta_i$ is the thickness of layer "$i$ and $r_{ij}$ is the reflection coefficient of the interface between layers "$i$" and "$j$"
\begin{equation}
r_{ij} = \frac{Q_i-Q_j}{Q_i+Q_j},  
\label{eq:r_ij}
\end{equation} 
where zero interface roughness is assumed. 

We note that any $|r_{ij}| \ll 1$ above the largest critical angle corresponding to the layer with the greatest refractive decrement $\delta$. 
In this approximation the net Fresnel reflection coefficient of a bi-layer (Fig.~\ref{fig:gzi}) is given by
\begin{equation}
r(Q) = r_{01} + r_{12}e^{iQ_1\Delta_1} + r_{23}e^{i(Q_1\Delta_1+Q_2\Delta_2)} + O (r_{ij}^3).
\label{eq:rq}
\end{equation}
The Fresnel transmissivity is given by
\begin{equation}
\begin{split}
T(Q) \simeq &  T_{01} + |r_{12}|^2 e^{-2Q"_1\Delta_1} + |r_{23}|^2 e^{-2(Q"_1\Delta_1 + Q"_2\Delta_2)} \\
& + r^*_{01}r_{12} e^{i Q_1 \Delta_1} + r^*_{12}r_{23} e^{i Q_2\Delta_2} e^{-2Q"_1\Delta_1} \\
& + r^*_{01}r_{23} e^{i(Q_1\Delta_1+Q_2\Delta_2)} + c.c.,
\end{split}
\label{eq:tt}
\end{equation}
where $T_{01} = 1 - |r_{01}|^2$ is the transmissivity of entrance interface of the thick mirror, the symbol "$^*$" denotes complex conjugation and "c.c." denotes conjugated components with the oscillating exponential factors. The reflection coefficients $r_{ij}$ are non-oscillating functions of the layer thicknesses (i.e., do not contain Kiessig fringes) \cite{Sakurai92}. If $\exp{[-2 Im(Q_1)\Delta_1]} \ll 1$ substantial attenuation of the transmitted wave occurs in the top layer (i.e., the second layer is deeply buried). Thus, according to Eq.~\ref{eq:tt} the Fresnel transmissivity of a bi-layer system will differ from that of the thick mirror by components oscillating with frequencies $\Delta_1$, $\Delta_2$, and the sum frequency $\Delta_1$ + $\Delta_2$. For the purpose of Fourier analysis in the ($Re(Q_i)$,$\Delta$)-space above the greatest critical angle one can assume $Re(Q_2) \approx Re(Q_1)$. Generalization of this analysis to the case of arbitrary number of layers is straightforward (see Supplemental Material). 

\begin{figure}[t]
\centering\includegraphics[width=0.45\textwidth]{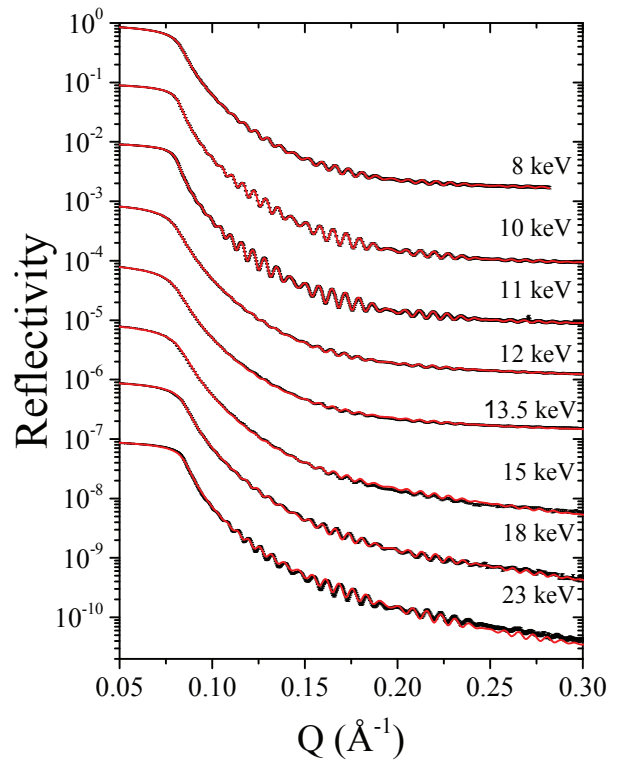}
\caption{Specular X-ray reflectivity (black squares) from the Pt/Cr bilayer deposited on the Si substrate, together with the fitting curves (solid red lines). The error bars indicate  $\pm$ 1 standard deviation. The X-ray energy at which the corresponding data were taken is indicated to the right. The curves are offset by one order of magnitude for clarity.}
\label{fig:refl}
\end{figure}

To address the problem experimentally integral photoemission yield and X-ray reflectivity of a bi-layer Pt-Cr X-ray mirror were measured simultaneously at several different photon energies in the range 8~-~23~keV. Prior to the analysis of integral electron yield the mirror was fully characterized using X-ray reflectivity (XRR), a method commonly used for structural characterization of surfaces \cite{Parratt54}.  As determined by XRR the thickness of the top Pt layer was $\Delta_1$~=~725.2~$\rm \AA$ and that of the buried Cr layer was $\Delta_2$~=~116.5~$\rm \AA$. The measured reflectivities and XRR fitting curves are shown in Fig.~\ref{fig:refl} (see Methods for details). 

\begin{figure*}[t]
\centering\includegraphics[width=0.9\textwidth]{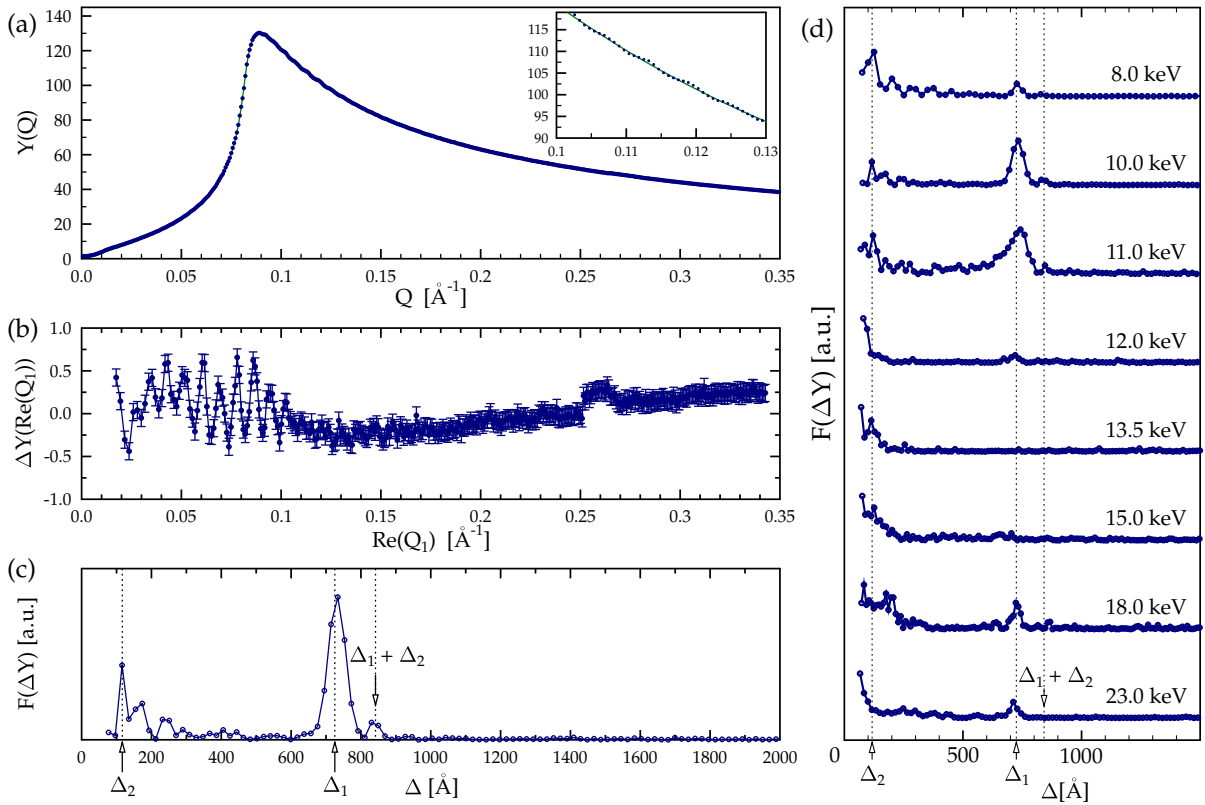}
\caption{Integral photoelectron yield and Fourier power spectra showing the structure of the Pt-Cr-Si bi-layer X-ray mirror. 
(a) Measured integral photoelectron yield as a function of the wavevector transfer $Q$ of the incident wave at 10~keV (circles) and fit to Eq.~\ref{eq:Y} corresponding to photoelectron yield of a thick Pt mirror (green solid line). The inset shows magnified region above the critical angle where the Kiessig fringes due to the layered structure of the mirror are clearly observed. (b) Differential photoelectron yield (circles, Y-error bars and solid line) obtained as the difference between the experimental data and the fit in (a). The differential yield is plotted as a function of the real part of $Q_1$, the wavevector transfer inside the Pt layer. (c) Forier power spectrum (circles, solid line) of the differential yield (b) showing peaks corresponding to thicknesses of individual layers ($\Delta_1$ and $\Delta_2$ and their sum. (d) Fourier power spectra (circles, solid lines) obtained at different photon energies (shifted for clarity).}
\label{fig:qy}
\end{figure*}

Figure~\ref{fig:qy} illustrates the strategy used to extract thicknesses of individual layers from the integral photoemission yield curve. In the first step the curves collected at different photon energies were renormalized in the units of quantum yield  (electric current divided by the incident flux and unit electric charge) and fit using Eq.~\ref{eq:Y}. The effective photoelectron escape depths were found to be $L \approx$ 200 - 400 $\rm \AA$ with greater values obtained at higher photon energies (see Supplemental Material for summary of the fit parameters). 
Figure~\ref{fig:qy}(a) shows the resulting experimental curve (blue circles) and the fit (solid green line) for the photon energy $E_X$~=~10~keV. The Kiessig fringes were clearly observed (magnified region shown in the inset of Fig.~\ref{fig:qy}(a)). In the next step, the fit was subtracted from the experimental data to isolate the differential yield containing the structural information. The differential yield $\Delta Y(Q_1)$ is shown in Fig.~\ref{fig:qy}(b) as a function of wavevector transfer $Re(Q_1)$ in the Pt layer. The power spectrum of a Fourier transform applied to $\Delta Y(Q_1)$ is shown in Fig.~\ref{fig:qy}(c). Peaks in the spectrum correspond to thicknesses of the individual layers with precision on the order of spectral resolution $2\pi/\Delta Q_1$~$\simeq$~10-18~$\rm \AA$ defined by the available range of 
the wavevector transfer $\Delta Q_1$.

The low-$\Delta$ region of the spectrum with peak corresponding to the buried Cr layer (116.5 $\AA$) could be contaminated by non-ideal subtraction procedure using the baseline, which includes the approximated photon-electron attenuation factor and the transmissivity of the thick mirror. Thus, evaluation of the thickness of the buried layer from the position of this peak alone can raise doubts. However, the appearance of the peak corresponding to the sum frequency $\Delta_1 +\Delta_2$ in the same spectrum confirms the presence of the Cr layer and its thickness. Also, the relatively featureless high-$\Delta$ region of the spectrum validates the approximation on the smallness of $|r_{ij}|$ given by Eq.~\ref{eq:rq}.  

Figure~\ref{fig:qy} shows power spectra of the differential yield of the bi-layer mirror at different photon energies. The peaks corresponding to individual thicknesses $\Delta_1$ and $\Delta_2$ as well as the sum $\Delta_1 + \Delta_2$ are clearly observed at photon energies 10~keV and 11~keV where Kiessig fringes are most intense. 
At some other photon energies (e.g., 13.5~keV and 15~keV) the noise level in the differential yield exceeds the weights of the spectral components. However, at higher photon energies (18~keV and 23~keV) reliable spectral detection of the thicknesses is still possible. Overall, the magnitude of the spectral components at various photon energies is consistent with the intensity of Kiessig fringes observed in XRR (Fig.~\ref{fig:refl}), which is governed by the energy-dependent absorption losses in the Pt layer. At 10-11~keV just below the energy region, which includes Pt L-absorption edges ($L_3$ at 11.56 keV, $L_2$ at 13.27 keV and $L_1$ at 13.88 keV) the imaginary parts of the refractive index ($\beta$) for Pt and Cr become comparable resulting in the increased intensity of the fringes. At the intermediate photon energies (12~keV - 15~keV) $\beta$ for Pt increases substantially in comparison to that of Cr. Finally, at higher photon energies (18~keV, 23~keV) the absorption losses for Pt become sufficiently small and the intensity of the fringes again increases. However, the absorption contrast between Pt and Cr remains substantial which explains reduction in the intensity of spectral peaks at $\Delta_2$ and $\Delta_1 + \Delta_2$. 

In summary, our findings demonstrate that structure of buried layers can be studied with hard X-rays, yet, without detection of the reflected radiation. Instead, integral photoelectron yield is detected using simple voltage-bias-driven collection of generated electric charges above the surface of the studied material.  
It should be mentioned that the approach is rather general and is not limited to the case where the top layer represents a conductive material. Indeed, self-detection of X-ray standing wave-effects have been observed on non-conductive materials such as diamond (e.g.,~\cite{Stoupin_SPIE14}). 
The approach could be used to advance the field of non-destructive evaluation where unique field applications can become feasible. In particular, it can be useful for quantitative evaluation of surfaces in enclosed geometries where implementation of the conventional $\theta - 2\theta$ XRR arrangement is not possible. Potentially, the use of very hard X-rays at very grazing incidence is feasible since contrary to XRR no effort is required to isolate the reflected radiation from the incident X-rays. Such approach could become indispensable for non-destructive evaluation of internal surfaces in nuclear reactors, hydrocarbon transport systems, internal combustion engines, buried transistors in semiconductor circuits and other systems with buried layers and interfaces.     

\section*{Methods}
\subsection*{Experiment}
\begin{figure}[t]
\centering\includegraphics[width=0.45\textwidth]{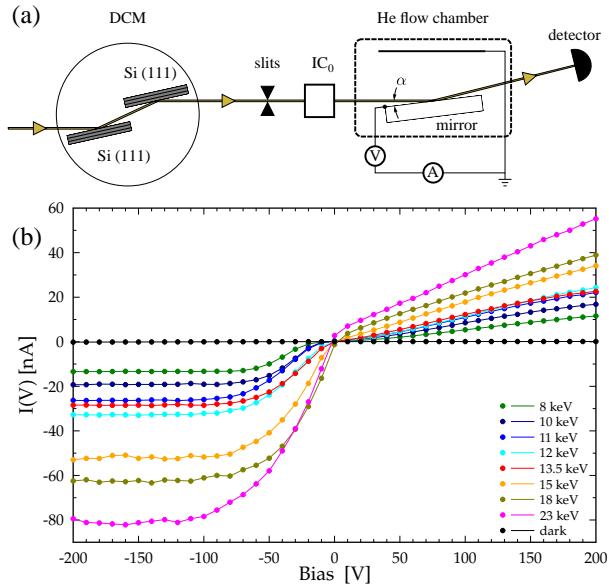}
\caption{(a) Experimental setup for simultaneous measurement of Fresnel reflectivity and the integral photoemission yield (see text for details). (b) IV-curves of the system at the angular peak of the photoemission yield measured at various photon energies.}
\label{fig:exper}
\end{figure} 

An X-ray mirror was prepared by deposition of a Pt film with thickness $\gtrsim 500 \mathrm{\AA}$ on top of a $\approx$~100-$\rm \AA$-thick layer of Cr on a polished Si substrate. The length of the mirror was 80~mm and the width was 20~mm. The mirror was placed on a non-conductive surface and contained in a helium flow chamber having an entrance and exit windows made of Kapton$^{\textregistered}$ film. The metallic working surface of the mirror served as the first electrode while a separate second grounded electrode was placed inside the chamber above the mirror at a distance of about 20 mm. The experiment was performed at 1-BM beamline of the Advanced Photon Source. The X-ray beam incident on the mirror was delivered by a Si (111) double-crystal monochromator (DCM). The DCM was detuned to suppress high-order Si Bragg reflections at high photon energies. The electric current between the electrodes in the flow chamber was measured using a source meter with applied bias voltages of $\pm$ 200 V. 
The observed maximum electric currents (at the photoemission peak) were $\simeq$~10~-~80 nA with greater values at higher photon energies. Electric current in the absence of x rays (i.e., dark current) was $\approx$~100~pA. The root mean square fluctuations in the measured signal were about 20~pA. A calibrated solid state detector was placed behind the flow chamber to measure the reflected X-ray flux (mirror in the beam) and the incident X-ray flux (mirror out of the beam).
The size of the incident beam was set to $0.1\times3.0$~mm$^2$ (vertical$\times$horizontal) using X-ray slits placed upstream of the mirror chamber. The incident beam was centered on the surface of the mirror. Simultaneous measurement of the reflectivity and the electric current were performed while scanning the mirror's grazing angle $\alpha$ at different photon energies selected by the double-crystal monochromator. Prior to each scan the incident photon flux was measured using the solid state detector. The measured values were in the range $7 \times 10^{8} - 2 \times 10^{9}$ photons/s. The experimental setup is shown in Fig.~\ref{fig:exper}(a). 

Prior to collection of angular curves an IV curve of the system was measured at the photoemission peak angle for each photon energy. The resulting IV curves are shown in Fig.~\ref{fig:exper}(b). In all cases saturation of the electric current and thus linear ionization-chamber-like response (full charge collection $\epsilon_q \simeq 1$) was achieved at sufficiently large negative potentials (-~200~V) applied to the mirror surface.  The absence of saturation at the positive potential (+~200~V ) suggests that the potential is insufficient to recapture all the electrons in the system. We note that in the case of Pd mirror studied earlier \cite{Stoupin_APL16} full charge collection was observed for both positive and negative potentials. Detailed explanation of this phenomenon requires consideration of energy spectra of the emitted photoelectrons and their energy transfer to He atoms via collision processes. For the purpose of linearity of detection of the integral photoelectron yield it was sufficient to ascertain the saturation at -~200~V in the present case. 

\subsection*{Structural analysis using XRR}
XRR measures the intensity of the specularly reflected beam as a function of the wavevector transfer $Q$ to extract the scattering length density (SLD) profile of the sample. SLD yields the information about the chemical composition of each layer, its thickness, physical density, and interfacial roughness. One of the main disadvantages of XRR is that spatial separation of the incident and the specularly reflected radiation becomes problematic in very grazing incidence ($\alpha \lesssim$~1~mrad) at high photon energies ($E_X \gtrsim$~20~keV). Reliable detection of the specularly reflected radiation requires $\theta$-$2\theta$ geometry and a large dynamic range of the radiation detector. These factors restrict the use of XRR to an X-ray analytical laboratory or a specialized X-ray source facility. Thus, despite the high penetrating power of hard X-rays, non-invasive structural studies of internal surfaces remain limited. 
In our case of a model system (bi-layer X-ray mirror) XRR was fully applicable. For the XRR modelling we used dynamic Parratt's formalism \cite{Parratt54} combined with Nevot--Croce interface roughness and genetic fitting algorithm using the MOTOFIT software \cite{Nelson06}. Literature SLD values for the Si substrate, Cr, and Pt layers were used for modelling and the individual thicknesses of Pt (705.2~$\rm \AA$) and Cr (116.5~$\rm \AA$) layers, as well as the interfacial roughness at Pt-Cr ($\sigma_{12}$~=~7.5~$\rm \AA$) and Cr-Si ($\sigma_{23}$~=~4.5~$\rm \AA$) interfaces, were kept the same for all energies. In this experiment, lateral projection of the coherence length $l_c$ (a few tens of microns) of the radiation is much smaller than the sample surface along the beam (80 mm). Therefore, the local surface roughness (within the length $l_c$)  from different parts of the sample is averaged incoherently. Such averaging is represented in the model by an additional top layer of ~20 $\rm \AA$-thick with SLD $\simeq$~80-85\% of the literature value for Pt. In other words, this top layer can be considered as a long-scale mirror height profile with a peak-to-valley value of 20 \AA, consistent with the Pt deposition specification.

\section*{Author contributions}
S.S. and M.Z. performed the experiments, analyzed the data and wrote the paper.
B.S. designed and fabricated the x-ray mirror. 

\section*{Acknowledgements}
K. Lang is acknowledged for technical support. Use of the Advanced Photon Source was supported by the U. S. Department of Energy, Office of Science, under Contract No. DE-AC02-06CH11357. The work at the National Synchrotron Light Source-II, Brookhaven National Laboratory, was supported by the U.S. Department of Energy, Office of Science, Office of Basic Energy Sciences, under Contract No. DE-SC0012704. 


\begin{thebibliography}{30}
\expandafter\ifx\csname natexlab\endcsname\relax\def\natexlab#1{#1}\fi
\expandafter\ifx\csname bibnamefont\endcsname\relax
  \def\bibnamefont#1{#1}\fi
\expandafter\ifx\csname bibfnamefont\endcsname\relax
  \def\bibfnamefont#1{#1}\fi
\expandafter\ifx\csname citenamefont\endcsname\relax
  \def\citenamefont#1{#1}\fi
\expandafter\ifx\csname url\endcsname\relax
  \def\url#1{\texttt{#1}}\fi
\expandafter\ifx\csname urlprefix\endcsname\relax\def\urlprefix{URL }\fi
\providecommand{\bibinfo}[2]{#2}
\providecommand{\eprint}[2][]{\url{#2}}

\bibitem[{\citenamefont{Henke}(1972)}]{Henke72}
\bibinfo{author}{\bibfnamefont{B.~L.} \bibnamefont{Henke}},
  \bibinfo{journal}{Phys. Rev. A} \textbf{\bibinfo{volume}{6}},
  \bibinfo{pages}{94} (\bibinfo{year}{1972}).

\bibitem[{\citenamefont{Fadley}(1974)}]{Fadley74}
\bibinfo{author}{\bibfnamefont{C.}~\bibnamefont{Fadley}}, \bibinfo{journal}{J.
  Electron Spectrosc. Relat. Phenom.} \textbf{\bibinfo{volume}{5}},
  \bibinfo{pages}{725 } (\bibinfo{year}{1974}).

\bibitem[{\citenamefont{Chester and Jach}(1993)}]{Chester93}
\bibinfo{author}{\bibfnamefont{M.~J.} \bibnamefont{Chester}} \bibnamefont{and}
  \bibinfo{author}{\bibfnamefont{T.}~\bibnamefont{Jach}},
  \bibinfo{journal}{Phys. Rev. B} \textbf{\bibinfo{volume}{48}},
  \bibinfo{pages}{17262} (\bibinfo{year}{1993}).

\bibitem[{\citenamefont{Kawai et~al.}(1995)\citenamefont{Kawai, Kawato,
  Hayashi, Horiuchi, Matsushige, and Kitajima}}]{Kawai95}
\bibinfo{author}{\bibfnamefont{J.}~\bibnamefont{Kawai}},
  \bibinfo{author}{\bibfnamefont{S.}~\bibnamefont{Kawato}},
  \bibinfo{author}{\bibfnamefont{K.}~\bibnamefont{Hayashi}},
  \bibinfo{author}{\bibfnamefont{T.}~\bibnamefont{Horiuchi}},
  \bibinfo{author}{\bibfnamefont{K.}~\bibnamefont{Matsushige}},
  \bibnamefont{and} \bibinfo{author}{\bibfnamefont{Y.}~\bibnamefont{Kitajima}},
  \bibinfo{journal}{Appl. Phys. Lett.} \textbf{\bibinfo{volume}{67}},
  \bibinfo{pages}{3889} (\bibinfo{year}{1995}).

\bibitem[{\citenamefont{Hayashi et~al.}(1996)\citenamefont{Hayashi, Kawato,
  Horiuchi, Matsushige, Kitajima, Takenaka, and Kawai}}]{Hayashi96}
\bibinfo{author}{\bibfnamefont{K.}~\bibnamefont{Hayashi}},
  \bibinfo{author}{\bibfnamefont{S.}~\bibnamefont{Kawato}},
  \bibinfo{author}{\bibfnamefont{T.}~\bibnamefont{Horiuchi}},
  \bibinfo{author}{\bibfnamefont{K.}~\bibnamefont{Matsushige}},
  \bibinfo{author}{\bibfnamefont{Y.}~\bibnamefont{Kitajima}},
  \bibinfo{author}{\bibfnamefont{H.}~\bibnamefont{Takenaka}}, \bibnamefont{and}
  \bibinfo{author}{\bibfnamefont{J.}~\bibnamefont{Kawai}},
  \bibinfo{journal}{Appl. Phys. Lett.} \textbf{\bibinfo{volume}{68}},
  \bibinfo{pages}{1921} (\bibinfo{year}{1996}).

\bibitem[{\citenamefont{Fadley}(2005)}]{Fadley05}
\bibinfo{author}{\bibfnamefont{C.~S.} \bibnamefont{Fadley}},
  \bibinfo{journal}{Nucl. Instrum. Methods Phys. Res. A}
  \textbf{\bibinfo{volume}{547}}, \bibinfo{pages}{24 } (\bibinfo{year}{2005}).

\bibitem[{\citenamefont{Fadley}(2010)}]{Fadley10}
\bibinfo{author}{\bibfnamefont{C.}~\bibnamefont{Fadley}}, \bibinfo{journal}{J.
  Electron Spectrosc. Relat. Phenom.} \textbf{\bibinfo{volume}{178–179}},
  \bibinfo{pages}{2 } (\bibinfo{year}{2010}).

\bibitem[{\citenamefont{Kawai}(2010)}]{Kawai10}
\bibinfo{author}{\bibfnamefont{J.}~\bibnamefont{Kawai}}, \bibinfo{journal}{J.
  Electron Spectrosc. Relat. Phenom.} \textbf{\bibinfo{volume}{178–179}},
  \bibinfo{pages}{268 } (\bibinfo{year}{2010}).

\bibitem[{\citenamefont{Sekiyama et~al.}(2000)\citenamefont{Sekiyama, Iwasaki,
  Matsuda, Saitoh, Onuki, and Suga}}]{Sekiyama00}
\bibinfo{author}{\bibfnamefont{A.}~\bibnamefont{Sekiyama}},
  \bibinfo{author}{\bibfnamefont{T.}~\bibnamefont{Iwasaki}},
  \bibinfo{author}{\bibfnamefont{K.}~\bibnamefont{Matsuda}},
  \bibinfo{author}{\bibfnamefont{Y.}~\bibnamefont{Saitoh}},
  \bibinfo{author}{\bibfnamefont{Y.}~\bibnamefont{Onuki}}, \bibnamefont{and}
  \bibinfo{author}{\bibfnamefont{S.}~\bibnamefont{Suga}},
  \bibinfo{journal}{Nature} \textbf{\bibinfo{volume}{403}},
  \bibinfo{pages}{396} (\bibinfo{year}{2000}).

\bibitem[{\citenamefont{Gray et~al.}(2012)\citenamefont{Gray, Minár, Ueda,
  Stone, Yamashita, Fujii, Braun, Plucinski, Schneider, Panaccione
  et~al.}}]{Gray12}
\bibinfo{author}{\bibfnamefont{A.~X.} \bibnamefont{Gray}},
  \bibinfo{author}{\bibfnamefont{J.}~\bibnamefont{Minár}},
  \bibinfo{author}{\bibfnamefont{S.}~\bibnamefont{Ueda}},
  \bibinfo{author}{\bibfnamefont{P.~R.} \bibnamefont{Stone}},
  \bibinfo{author}{\bibfnamefont{Y.}~\bibnamefont{Yamashita}},
  \bibinfo{author}{\bibfnamefont{J.}~\bibnamefont{Fujii}},
  \bibinfo{author}{\bibfnamefont{J.}~\bibnamefont{Braun}},
  \bibinfo{author}{\bibfnamefont{L.}~\bibnamefont{Plucinski}},
  \bibinfo{author}{\bibfnamefont{C.~M.} \bibnamefont{Schneider}},
  \bibinfo{author}{\bibfnamefont{G.}~\bibnamefont{Panaccione}},
  \bibnamefont{et~al.}, \bibinfo{journal}{Nat. Mater.}
  \textbf{\bibinfo{volume}{11}}, \bibinfo{pages}{957} (\bibinfo{year}{2012}).

\bibitem[{\citenamefont{Feng}(2011)}]{DLFeng11}
\bibinfo{author}{\bibfnamefont{D.-L.} \bibnamefont{Feng}},
  \bibinfo{journal}{Nat. Mater.} \textbf{\bibinfo{volume}{10}},
  \bibinfo{pages}{729} (\bibinfo{year}{2011}).

\bibitem[{\citenamefont{Dallera et~al.}(2004)\citenamefont{Dallera, Duò,
  Braicovich, Panaccione, Paolicelli, Cowie, and Zegenhagen}}]{Dallera04}
\bibinfo{author}{\bibfnamefont{C.}~\bibnamefont{Dallera}},
  \bibinfo{author}{\bibfnamefont{L.}~\bibnamefont{Duò}},
  \bibinfo{author}{\bibfnamefont{L.}~\bibnamefont{Braicovich}},
  \bibinfo{author}{\bibfnamefont{G.}~\bibnamefont{Panaccione}},
  \bibinfo{author}{\bibfnamefont{G.}~\bibnamefont{Paolicelli}},
  \bibinfo{author}{\bibfnamefont{B.}~\bibnamefont{Cowie}}, \bibnamefont{and}
  \bibinfo{author}{\bibfnamefont{J.}~\bibnamefont{Zegenhagen}},
  \bibinfo{journal}{Appl. Phys. Lett.} \textbf{\bibinfo{volume}{85}}
  (\bibinfo{year}{2004}).

\bibitem[{\citenamefont{Martens et~al.}(1978)\citenamefont{Martens, Rabe,
  Schwentner, and Werner}}]{Martens78}
\bibinfo{author}{\bibfnamefont{G.}~\bibnamefont{Martens}},
  \bibinfo{author}{\bibfnamefont{P.}~\bibnamefont{Rabe}},
  \bibinfo{author}{\bibfnamefont{N.}~\bibnamefont{Schwentner}},
  \bibnamefont{and} \bibinfo{author}{\bibfnamefont{A.}~\bibnamefont{Werner}},
  \bibinfo{journal}{J. Phys. C: Solid St. Phys.} \textbf{\bibinfo{volume}{11}},
  \bibinfo{pages}{3125} (\bibinfo{year}{1978}).

\bibitem[{\citenamefont{Martens et~al.}(1979)\citenamefont{Martens, Rabe,
  Tolkiehn, and Werner}}]{Martens79}
\bibinfo{author}{\bibfnamefont{G.}~\bibnamefont{Martens}},
  \bibinfo{author}{\bibfnamefont{P.}~\bibnamefont{Rabe}},
  \bibinfo{author}{\bibfnamefont{G.}~\bibnamefont{Tolkiehn}}, \bibnamefont{and}
  \bibinfo{author}{\bibfnamefont{A.}~\bibnamefont{Werner}},
  \bibinfo{journal}{Phys. Stat. Solidi (a)} \textbf{\bibinfo{volume}{55}},
  \bibinfo{pages}{105} (\bibinfo{year}{1979}).

\bibitem[{\citenamefont{Erbil et~al.}(1988)\citenamefont{Erbil, Cargill~III,
  Frahm, and Boehme}}]{Erbil88}
\bibinfo{author}{\bibfnamefont{A.}~\bibnamefont{Erbil}},
  \bibinfo{author}{\bibfnamefont{G.~S.} \bibnamefont{Cargill~III}},
  \bibinfo{author}{\bibfnamefont{R.}~\bibnamefont{Frahm}}, \bibnamefont{and}
  \bibinfo{author}{\bibfnamefont{R.~F.} \bibnamefont{Boehme}},
  \bibinfo{journal}{Phys. Rev. B} \textbf{\bibinfo{volume}{37}},
  \bibinfo{pages}{2450} (\bibinfo{year}{1988}).

\bibitem[{\citenamefont{Kordesch and Hoffman}(1984)}]{Kordesch84}
\bibinfo{author}{\bibfnamefont{M.~E.} \bibnamefont{Kordesch}} \bibnamefont{and}
  \bibinfo{author}{\bibfnamefont{R.~W.} \bibnamefont{Hoffman}},
  \bibinfo{journal}{Phys. Rev. B} \textbf{\bibinfo{volume}{29}},
  \bibinfo{pages}{491} (\bibinfo{year}{1984}).

\bibitem[{\citenamefont{Guo and denBoer}(1985)}]{Guo85}
\bibinfo{author}{\bibfnamefont{T.}~\bibnamefont{Guo}} \bibnamefont{and}
  \bibinfo{author}{\bibfnamefont{M.~L.} \bibnamefont{denBoer}},
  \bibinfo{journal}{Phys. Rev. B} \textbf{\bibinfo{volume}{31}},
  \bibinfo{pages}{6233} (\bibinfo{year}{1985}).

\bibitem[{\citenamefont{Jones et~al.}(1978)\citenamefont{Jones, Thomas, Thorpe,
  and Tricker}}]{Jones78}
\bibinfo{author}{\bibfnamefont{W.}~\bibnamefont{Jones}},
  \bibinfo{author}{\bibfnamefont{J.~M.} \bibnamefont{Thomas}},
  \bibinfo{author}{\bibfnamefont{R.~K.} \bibnamefont{Thorpe}},
  \bibnamefont{and} \bibinfo{author}{\bibfnamefont{M.~J.}
  \bibnamefont{Tricker}}, \bibinfo{journal}{Applications of Surface Science}
  \textbf{\bibinfo{volume}{1}}, \bibinfo{pages}{388 } (\bibinfo{year}{1978}).

\bibitem[{\citenamefont{Stoupin}(2016)}]{Stoupin_APL16}
\bibinfo{author}{\bibfnamefont{S.}~\bibnamefont{Stoupin}},
  \bibinfo{journal}{Appl. Phys. Lett.} \textbf{\bibinfo{volume}{108}},
  \bibinfo{eid}{041101} (\bibinfo{year}{2016}).

\bibitem[{\citenamefont{St{\"o}hr}(1992)}]{Stohr_book}
\bibinfo{author}{\bibfnamefont{J.}~\bibnamefont{St{\"o}hr}},
  \emph{\bibinfo{title}{NEXAFS spectroscopy}}, vol.~\bibinfo{volume}{25} of
  \emph{\bibinfo{series}{Springer Series in Surface Sciences}}
  (\bibinfo{publisher}{Springer}, \bibinfo{address}{Berlin Heidelberg
  New~York}, \bibinfo{year}{1992}).

\bibitem[{\citenamefont{Weiss and Bernstein}(1956)}]{Weiss56}
\bibinfo{author}{\bibfnamefont{J.}~\bibnamefont{Weiss}} \bibnamefont{and}
  \bibinfo{author}{\bibfnamefont{W.}~\bibnamefont{Bernstein}},
  \bibinfo{journal}{Phys. Rev.} \textbf{\bibinfo{volume}{103}},
  \bibinfo{pages}{1253} (\bibinfo{year}{1956}).

\bibitem[{\citenamefont{Henke et~al.}(1977)\citenamefont{Henke, Smith, and
  Attwood}}]{Henke77}
\bibinfo{author}{\bibfnamefont{B.~L.} \bibnamefont{Henke}},
  \bibinfo{author}{\bibfnamefont{J.~A.} \bibnamefont{Smith}}, \bibnamefont{and}
  \bibinfo{author}{\bibfnamefont{D.~T.} \bibnamefont{Attwood}},
  \bibinfo{journal}{J. Appl. Phys.} \textbf{\bibinfo{volume}{48}},
  \bibinfo{pages}{1852} (\bibinfo{year}{1977}).

\bibitem[{\citenamefont{Shah et~al.}(1988)\citenamefont{Shah, Elliott,
  McCallion, and Gilbody}}]{Shah88}
\bibinfo{author}{\bibfnamefont{M.~B.} \bibnamefont{Shah}},
  \bibinfo{author}{\bibfnamefont{D.~S.} \bibnamefont{Elliott}},
  \bibinfo{author}{\bibfnamefont{P.}~\bibnamefont{McCallion}},
  \bibnamefont{and} \bibinfo{author}{\bibfnamefont{H.~B.}
  \bibnamefont{Gilbody}}, \bibinfo{journal}{J. Phys. B: At. Mol. Opt. Phys}
  \textbf{\bibinfo{volume}{21}}, \bibinfo{pages}{2751} (\bibinfo{year}{1988}).

\bibitem[{\citenamefont{Yang et~al.}(2013)\citenamefont{Yang, Gray, Kaiser,
  Mun, Sell, Kortright, and Fadley}}]{Yang13}
\bibinfo{author}{\bibfnamefont{S.-H.} \bibnamefont{Yang}},
  \bibinfo{author}{\bibfnamefont{A.~X.} \bibnamefont{Gray}},
  \bibinfo{author}{\bibfnamefont{A.~M.} \bibnamefont{Kaiser}},
  \bibinfo{author}{\bibfnamefont{B.~S.} \bibnamefont{Mun}},
  \bibinfo{author}{\bibfnamefont{B.~C.} \bibnamefont{Sell}},
  \bibinfo{author}{\bibfnamefont{J.~B.} \bibnamefont{Kortright}},
  \bibnamefont{and} \bibinfo{author}{\bibfnamefont{C.~S.}
  \bibnamefont{Fadley}}, \bibinfo{journal}{J. Appl. Phys.}
  \textbf{\bibinfo{volume}{113}}, \bibinfo{eid}{073513} (\bibinfo{year}{2013}).

\bibitem[{\citenamefont{Sakurai and Iida}(1992)}]{Sakurai92}
\bibinfo{author}{\bibfnamefont{K.}~\bibnamefont{Sakurai}} \bibnamefont{and}
  \bibinfo{author}{\bibfnamefont{A.}~\bibnamefont{Iida}},
  \bibinfo{journal}{Japanese Journal of Applied Physics}
  \textbf{\bibinfo{volume}{31}}, \bibinfo{pages}{L113} (\bibinfo{year}{1992}).

\bibitem[{\citenamefont{Parratt}(1954)}]{Parratt54}
\bibinfo{author}{\bibfnamefont{L.~G.} \bibnamefont{Parratt}},
  \bibinfo{journal}{Phys. Rev.} \textbf{\bibinfo{volume}{95}},
  \bibinfo{pages}{359} (\bibinfo{year}{1954}).

\bibitem[{\citenamefont{Hau-Riege}(2011)}]{Hau-Riege_book}
\bibinfo{author}{\bibfnamefont{S.~P.} \bibnamefont{Hau-Riege}},
  \emph{\bibinfo{title}{High-Intensity X-Rays – Interaction with Matter}}
  (\bibinfo{publisher}{WILEY-VCH Verlag GmbH\&Co. KGaA},
  \bibinfo{address}{Boschstr. 12, 69469, Weinheim, Germany},
  \bibinfo{year}{2011}).

\bibitem[{\citenamefont{Windt}(1998)}]{Windt98}
\bibinfo{author}{\bibfnamefont{D.~L.} \bibnamefont{Windt}},
  \bibinfo{journal}{Computers in Physics} \textbf{\bibinfo{volume}{12}},
  \bibinfo{pages}{360} (\bibinfo{year}{1998}).

\bibitem[{\citenamefont{Stoupin et~al.}(2014)\citenamefont{Stoupin, Baryshev,
  and Antipov}}]{Stoupin_SPIE14}
\bibinfo{author}{\bibfnamefont{S.}~\bibnamefont{Stoupin}},
  \bibinfo{author}{\bibfnamefont{S.~V.} \bibnamefont{Baryshev}},
  \bibnamefont{and} \bibinfo{author}{\bibfnamefont{S.~P.}
  \bibnamefont{Antipov}}, \bibinfo{journal}{Proc. SPIE - Int. Soc. Opt. Eng.}
  \textbf{\bibinfo{volume}{9207}} (\bibinfo{year}{2014}), \bibinfo{note}{doi:
  10.1117/12.2062495}.

\bibitem[{\citenamefont{Nelson}(2006)}]{Nelson06}
\bibinfo{author}{\bibfnamefont{A.}~\bibnamefont{Nelson}}, \bibinfo{journal}{J.
  Appl. Cryst.} \textbf{\bibinfo{volume}{39}}, \bibinfo{pages}{273}
  (\bibinfo{year}{2006}).

\end{thebibliography}

\end{document}